\documentclass[english]{aastex6}
\usepackage[T1]{fontenc}
\usepackage{color}
\usepackage{graphicx}

\makeatletter
\slugcomment{}
\shorttitle{}
\shortauthors{}

\makeatother

\usepackage{babel}
\begin{document}

\title{Near-Infrared Imaging Polarimetry of Inner Region of GG Tau A
Disk}

\author{
Yi YANG\altaffilmark{1,2},
Jun HASHIMOTO\altaffilmark{3},
Saeko S. HAYASHI\altaffilmark{1,4},
Motohide TAMURA\altaffilmark{5,3,2},
Satoshi MAYAMA\altaffilmark{1,2},
Roman RAFIKOV\altaffilmark{25},
Eiji AKIYAMA\altaffilmark{2},
Joseph C. CARSON\altaffilmark{9},
Markus JANSON\altaffilmark{28},
Jungmi KWON\altaffilmark{10},
Jerome DE LEON\altaffilmark{5},
Daehyeon OH\altaffilmark{27,1,2},
Michihiro TAKAMI\altaffilmark{20},
Ya-wen TANG\altaffilmark{20},
Tomoyuki KUDO\altaffilmark{4},
Nobuhiko KUSAKABE\altaffilmark{3},
Lyu ABE\altaffilmark{6},
Wolfgang BRANDNER\altaffilmark{7},
Timothy D. BRANDT\altaffilmark{25},
Sebastian EGNER\altaffilmark{29},
Markus FELDT\altaffilmark{7},
Miwa GOTO\altaffilmark{11},
Carol A. GRADY\altaffilmark{12,13,14},
Olivier GUYON\altaffilmark{4},
Yutaka HAYANO\altaffilmark{1,2,4},
Masahiko HAYASHI\altaffilmark{1,2}
Thomas HENNING\altaffilmark{7},
Klaus W. HODAPP\altaffilmark{15},
Miki ISHII\altaffilmark{2},
Masanori IYE\altaffilmark{2},
Ryo KANDORI\altaffilmark{2},
Gillian R. KNAPP\altaffilmark{8},
Masayuki KUZUHARA \altaffilmark{3},
Taro MATSUO\altaffilmark{16},
Michael W. MCELWAIN\altaffilmark{12},
Shoken MIYAMA\altaffilmark{17},
Jun-Ichi MORINO\altaffilmark{2},
Amaya MORO-MARTIN\altaffilmark{8,18},
Tetsuo NISHIMURA\altaffilmark{4},
Tae-Soo PYO\altaffilmark{1,4},
Eugene SERABYN\altaffilmark{19},
Takuya SUENAGA\altaffilmark{1,2},
Hiroshi SUTO\altaffilmark{2,3},
Ryuji SUZUKI\altaffilmark{2},
Yasuhiro H. TAKAHASHI\altaffilmark{2,5},
Naruhisa TAKATO\altaffilmark{1,4},
Hiroshi TERADA\altaffilmark{2},
Christian THALMANN\altaffilmark{21},
Edwin L. TURNER\altaffilmark{8,22},
Makoto WATANABE\altaffilmark{23},
John WISNIEWSKI\altaffilmark{26},
Toru YAMADA\altaffilmark{24},
Hideki TAKAMI\altaffilmark{1,2},
Tomonori USUDA\altaffilmark{1,2}
}

\affil{
1 Department of Astronomical Science, The Graduate University for Advanced Studies, 2-21-1 Osawa, Mitaka, Tokyo 181-8588, Japan\\
2 National Astronomical Observatory of Japan (NAOJ), National Institutes of Natural Sciences (NINS), 2-21-1 Osawa, Mitaka, Tokyo 181-8588, Japan\\
3 Astrobiology Center, NINS, 2-21-1, Osawa, Mitaka, Tokyo, 181-8588, Japan\\
4 Subaru Telescope, NAOJ, NINS, 650 North A'ohoku Place, Hilo, HI 96720, USA\\
5 Department of Astronomy, The University of Tokyo, 7-3-1 Hongo, Bunkyo-ku, Tokyo 113-0033, Japan\\
6 Laboratoire Lagrange (UMR 7293), Universite de Nice-Sophia Antipolis, CNRS, Observatoire de la Coted'azur, 28 avenue Valrose, 06108 Nice Cedex 2, France\\
7 Max Planck Institute for Astronomy, K{\"o}nigstuhl 17, 69117 Heidelberg, Germany\\
8 Department of Astrophysical Science, Princeton University, Peyton Hall, Ivy Lane, Princeton, NJ 08544, USA\\
9 Department of Physics and Astronomy, College of Charleston, 58 Coming St., Charleston, SC 29424, USA\\
10 Institute of Space and Astronautical Science, Japan Aerospace Exploration Agency, 3-1-1 Yoshinodai, Chuo-ku, Sagamihara, Kanagawa 252-5210, Japan\\
11 Universit{\"a}ts-Sternwarte M{\"u}nchen, Ludwig-Maximilians-Universit{\"a}t, Scheinerstr. 1, 81679 M{\"u}nchen, Germany\\
12 Exoplanets and Stellar Astrophysics Laboratory, Code 667, Goddard Space Flight Center, Greenbelt, MD\\
20771, USA\\
13 Eureka Scientific, 2452 Delmer, Suite 100, Oakland CA 96002, USA\\
14 Goddard Center for Astrobiology\\
15 Institute for Astronomy, University of Hawaii, 640 N. A'ohoku Place, Hilo, HI 96720, USA\\
16 Department of Earth and Space Science, Graduate School of Science, Osaka University, 1-1 Machikaneyamacho, Toyonaka, Osaka 560-0043, Japan\\
17 Hiroshima University, 1-3-2 Kagamiyama, Higashihiroshima, Hiroshima 739-8511, Japan\\
18 Department of Astrophysics, CAB-CSIC/INTA, 28850 Torrej\'on de Ardoz, Madrid, Spain\\
19 Jet Propulsion Laboratory, California Institute of Technology, M/S 171-113 4800 Oak Grove Drive Pasadena, CA 91109 USA\\
20 Institute of Astronomy and Astrophysics, Academia Sinica, P.O. Box 23-141, Taipei 10617, Taiwan\\
21 Swiss Federal Institute of Technology (ETH Zurich), Institute for Astronomy,\\
Wolfgang-Pauli-Strasse 27, CH-8093 Zurich, Switzerland\\
22 Kavli Institute for Physics and Mathematics of the Universe, The University of Tokyo, 5-1-5 Kashiwanoha, Kashiwa, Chiba 277-8568, Japan\\
23 Department of Cosmosciences, Hokkaido University, Kita-ku, Sapporo, Hokkaido 060-0810, Japan\\
24 Astronomical Institute, Tohoku University, Aoba-ku, Sendai, Miyagi 980-8578, Japan\\
25 Astrophysics Department, Institute for Advanced Study, Princeton, NJ 08540, USA\\
26 H. L. Dodge Department of Physics \& Astronomy, University of Oklahoma, 440 W Brooks St., Norman, OK 73019, USA\\
27 National Meteorological Satellite Center, 64-18, Guam-gil, Gwanghyewon-myeon, Jincheon-gun, Chungbuk, South Korea\\
28 Department of Astronomy, Stockholm University, AlbaNova University Center, SE-106 91 Stockholm, Sweden\\
29 European Southern Observatory, Karl Schwarzschildstr. 2, Garching 85748, Germany\\
}

\email{yi.yang@nao.ac.jp}

\begin{abstract}

By performing non-masked polarization imaging with Subaru/HiCIAO,
polarized scattered light from the inner region of the disk around the GG Tau A system was successfully detected in the $H$ band with a spatial resolution of approximately 0.07$\arcsec$, revealing the complicated inner disk structures around this young binary.
This paper reports the observation of an arc-like structure to the north of GG Tau Ab
and part of a circumstellar structure that 
is noticeable 
around GG Tau Aa extending to a distance of approximately 28 AU from the primary star. The speckle noise around GG Tau Ab constrains its disk radius
to <13 AU. Based on the size of the circumbinary ring and the circumstellar disk around GG Tau Aa, the semi-major axis of the binary's orbit is likely to be 62 AU. A comparison of the present observations with previous ALMA and near-infrared (NIR) H$_2$ emission observations suggests that the north
arc could be part of a large streamer flowing from the circumbinary ring to sustain the circumstellar disks. According to the previous studies, the circumstellar disk around GG Tau Aa has enough mass and can 
sustain itself
for a duration sufficient for planet formation; thus, our study indicates that planets can form within close (separation $\lesssim$ 100 AU) young binary systems.

\end{abstract}


\section{Introduction}

Many stars in our galaxy form binary or multiple systems. \citet{Duchene2013}
noted that for solar-type ($0.7M_{\odot}$--$1.3M_{\odot}$) main sequence
stars, the multiple frequency can reach 44\%. Furthermore, many planets 
have been found in binary or multiple systems
such as $\tau$ Bo{\"o}tis Ab \citep{Butler1997},
Kepler-16 (AB) b \citep{Doyle2011}, and ROXs 42Bb \citep{Currie2014}.

\citet{Roell2012} estimated that at least 12\% of planet-hosting stars may be
binary or multiple systems, whereas \citet{Raghavan2006}
estimated this value to be 23\%, and \citet{Horch2014} very optimistically estimated
that approximately $40$--$50\%$ planet-hosting stars are binary stars. 
This begs the question of how planets form and evolve in binary or multiple systems.
In addition, the proportion of young stars in binary or multiple systems
appears to be 
two times higher
that for solar-type field stars \citep{Duchene2013}.
This may imply that stars tend to form from binary or multiple systems.
Furthermore, this indicates that even planets discovered around single stars may
have been affected by companion stars during their formation and evolution.
Therefore, to understand the early stages of the planet formation process,
research on planets and disks around binary or multiple systems is necessary.

Previous studies (e.g., \citet{Wang2015}) have demonstrated that the efficiency of planet formation in wide-separation ($\gtrsim$100 AU) binaries is not very different from their single star analogs. 
On the other hand,
disks in smaller-separation binaries ($\lesssim$100 AU) may be too disturbed by the companion's gravity and too short-lived to produce planets \citep{Duchene2010}.
Despite this fact, some planets have been discovered in close binaries. For example, S-type planets, which are planets that orbit around one of the binary stars, have been found in binaries with separations of approximately 20 AU (e.g., $\gamma$ Cep Ab \citep{Hatzes2003}),
and P-type planets, which are planets that orbit around both binary stars, have been found in binaries with separations of approximately 0.22 AU (e.g., Kepler-16 (AB)-b \citep{Doyle2011}).
In a census of a star formation region, \citet{Kraus2012b}
determined that although approximately 2/3 of the close binaries with a separation of $\lesssim$ 40 AU
lose their disks within approximately 1 Myr, the remainder of 
approximately 1/3 of the disks appear to experience an evolutionary timescale
similar to that of disks around single stars; thus, planets may have opportunities to form in binary systems. 
These results indicate that some mechanism may help planets form in 
such close
binaries. 
Some theories suggest that an additional star might have helped the planet formation process, 
such as changing the orbit of a planet through the Kozai--Lidov mechanism \citep{Kozai1962,Lidov1962}, causing the protoplanetary disks to become eccentric by truncating them (e.g., \citealp{Regaly2011}), or opening large gaps in circumbinary disks (e.g., \citealp{Artymowicz1994}). 
Thus, the planet formation process around binaries could be quite different from and more complicated than that around single stars. 
To obtain a better understanding of this, it is necessary to investigate disk structures in binaries to determine how the disks in binary systems evolve.

GG Tau is a well-known young multiple star system in the Taurus--Auriga molecular cloud at a distance of approximately 140 pc from the Solar System \citep{Kenyon1994}.  
This is a double binary system: 
GG Tau Aa/Ab and GG Tau Ba/Bb. GG Tau A is an eccentric ($e\simeq0.35$, \citet{Beust2005}) T Tauri binary system with an age of approximately 2.3 Myr \citep{Palla2002}. 
It consists of GG Tau Aa ($0.73M_{\odot}$) and GG Tau Ab ($0.64M_{\odot}$) with a separation of approximately 0.25$\arcsec$ (35 AU; \citealp{Kraus2009}). 
In addition, \citet{DiFolco2014} reported that GG Tau Ab is a binary with a separation of approximately 0.03$\arcsec$ (4.2 AU) and a period of approximately 16 yr; thus, this system is
actually a triple system, though it may still be regarded as a binary system in observations with 0.07$\arcsec$ resolution.

The GG Tau A system is noteworthy for its circumbinary ring, which was first discovered
by ground-based adaptive optics (AO) imaging \citep{Roddier1996} and
has been observed many times in various wavelengths, 
e.g., \citet{Guilloteau1999}
in the millimeter band, \citet{Krist2005} in the optical band, and \citet{Itoh2014}
in the NIR band. 
Millimeter and submillimeter observations \citep{Guilloteau1999,Dutrey2014} have shown that this ring rotates
clockwise and the northern edge is nearest to us. Additionally, a gap has been observed
in the northwestern region of the ring; e.g., \citet{Silber2000} observed it in 1998
with the Near Infrared Camera and Multi-Object Spectrometer installed on the Hubble Space Telescope (HST/NICMOS), and \citet{Krist2002} observed it in 1997 with the Wide Field and Planetary Camera 2 installed on the Hubble Space Telescope (HST/WFPC2).
This gap is believed to be a shadow cast by circumstellar materials \citep{Krist2005,Itoh2014}.

Several groups have investigated the disk structure inside the circumbinary ring of GG Tau A.
\citet{Pietu2011} suggested the possible ($2\sigma$) existence of a streamer
extending from the northeastern edge of the outer ring to the inner disk 
based on the observations from
the IRAM Plateau de Bure interferometer in a 1.1 mm continuum band.  
\citet{Beck2012} observed the H$_2 \
\nu=1-0 \  S(1)$ emission around the stars using the Gemini North Telescope,
arguing
that the strong emissions around the disk are likely caused
not by X-ray excitation but by shock waves due to an accretion flow in the disk.
\citet{Dutrey2014} detected a feature in their observations of the gas in CO J=6--5 transition
by the Atacama Large Millimeter Array (ALMA) implying the presence of streamers and 
speculated that the streamer may feed material from the outer region of the disk to a planet, sustaining planet formation.
These studies strongly indicate that the region inside the circumbinary disk may not have been cleared yet, 
but they have not revealed the detailed structure of this region because of their low spatial resolution. 
Investigating the details inside the circumbinary ring will be quite helpful in improving our understanding
of planet formation in this system; thus, it is very important to observe this region around
GG Tau A with a higher spatial resolution in the NIR band. 
In the past studies,
high-spatial-resolution NIR observations have helped to reveal the
structures around the binary system SR24 \citep{Mayama2010}, and it is thus
a promising method for improving our understanding of disk structures.

This paper discusses the successful
observation of the detailed structures
inside the circumbinary ring around GG Tau A,
which shows a "north arc" structure in the H-band that is believed to be part of a streamer flowing
from the circumbinary ring to GG Tau Ab. 
In Section 2, the observation
and data reduction processes are introduced. Section 3 presents the observation
results of GG Tau A. In Section 4, we compare the observations made in this study 
with theory and previous observations.
 Conclusions are given in Section 5.

\section{Observations and Data Reduction}

The presently reported observation of GG Tau A was performed on 8 January 2015 Hawaii
Standard Time using the Subaru 8.2 m Telescope with the High Contrast
Instrument for the Subaru Next Generation Adaptive Optics (HiCIAO;
\citealt{Tamura2006}) and the adaptive optics (AO) instrument AO188 \citep{Hayano2010}.
This observation was part of the survey program Strategic Explorations of Exoplanets
and Disks with Subaru (SEEDS), which began in 2009. 
This observation employed the quad-polarized differential imaging (qPDI) mode, which
uses a double-Wollaston prism to split the light into four $512\times512$
channels on the detector with pixel scale of 9.50 mas/pixel. 
To help reduce the saturated radius, two of these four channels each corresponded to o- and e-polarizations. During this observation, the AO system
limited the full width at half maximum of the stellar point spread function to 0.07$\arcsec$,
which is close to the diffraction limit of 0.04$\arcsec$. In a previous observation of GG Tau A 
by \citet{Itoh2014} in 2011, a mask with a 0.6$\arcsec$
diameter was used to obscure structures near the stars. To
help reveal the inner region of the disk, no mask was used
in the present observation. A half-wave
plate was used in the observation, and it was rotated among position
angles of $0^{\circ}$, $22.5^{\circ}$, $45^{\circ}$, and $67.5^{\circ}$
to measure the Stokes parameters. This cycle was repeated 15 times during observation. Ultimately, 60 frames were collected, each with
an exposure time of 5 s and 4 coadds. The
total integration time was 20 minutes.

The data reduction process was completed using the Image Reduction
and Analysis Facility (IRAF) pipeline. Flat field was corrected, and stripes, bad pixels, and distortions were removed. After these steps, the images were first cross-correlated in different channels. The Stokes parameters $+Q$, $+U$, $-Q$, and $-U$ were then obtained by subtracting the e-images from the o-images. Next, these Stokes images were aligned, and the $Q$ and $U$ images were constructed as $Q=((+Q)-(-Q))/2$, $U=((+U)-(-U))/2$. The Stokes $I$ image, or intensity image, was derived by averaging the sum of the o- and e- images in all frames. After the instrumental polarization was corrected,
a polarized intensity ($PI$) image was constructed as
$PI=\sqrt{U^{2}+Q^{2}}$. The uncorrected point spread function (PSF) halo
can disturb the polarization vectors; thus, measures were taken to remove
it (see Appendix for details). Because the PSF reference star was not obtained during the observation, it was difficult to remove the PSF from the Stokes $I$ image, which is actually a mixture of the total intensity image of the disk and the much brighter PSF of the binary. Therefore, the total intensity image as well as the polarization degree image ($P$ image) of the disk could not be derived, and the present discussion will be mainly based on the $PI$ image rather than the $P$ image. 

\section{Results}

The $PI$ image tracing the light scattered
by the dust grains of GG Tau A is shown in Figure 1, which provides a
wide view of GG Tau A and its disk. The field of view is $512\times512$ pixels, corresponding
to approximately $4.9\arcsec\times4.9\arcsec$. 
This image shows the circumbinary
ring, the two companion stars GG Tau Aa/Ab and some disk structures near them. 
Polarimetry observation is a powerful method of revealing circumstellar disk structures because it traces the polarized light scattered from the disks.
The nonpolarized light from the central stars is subtracted during data reduction. 
However, it should also be noted that a lack of polarized light does not necessarily mean that there is no scattered light or scattering structures (e.g., \citet{Perrin2009}). 
For the present discussion, the part of the disk inside the circumbinary ring
is defined as the "inner region" of the circumbinary disk around GG Tau A.

The separation between the two companion stars was derived as 0.27$\pm0.01\arcsec$, which corresponds
to 38$\pm$1 AU\footnote{In this paper, the sizes in AU are calculated by assuming that GG Tau A lies at a distance of 140 pc from the Solar System.}.
The position angle (PA) of GG Tau Aa/Ab binary is $327\pm1^{\circ}$ (measured from north to east).
The GG Tau Ab binary reported by \citet{DiFolco2014} could not be resolved. 

In the $PI$ image, the circumbinary ring looks asymmetric.
There seem to be offsets among the center of the outer edge ellipse, that of the inner edge ellipse, and the barycenter of the binary. 
To estimate the basic parameters of the circumbinary ring, we developed a toy model. 
In this model, we assumed the outer edge of the ring to be circular, the inclination $37^{\circ}$, and the PA i$277^{\circ}$, as in the previous observations. 
In addition, we attributed the barycenter of the binary to be near one of the foci of the inner edge ellipse. 
The result is shown in Figure 1(b). 
It was determined that the inner edge can be generally fitted by an ellipse with an eccentricity of approximately 0.2 
and there are offsets among the center of the inner edge ellipse, that of the outer edge ellipse, and the barycenter of the GG Tau A binary.
The center of the inner edge (yellow cross) is located approximately 15 pixels (0.14$\arcsec$ or 20 AU) to the south of the barycenter, 
which should be near one of the foci of the inner edge (yellow star), 
and that of the outer ring (red cross) is approximately 25 pixels (0.24$\arcsec$ or 33 AU) to the south of the center of the inner edge. 
This reveals the asymmetric characteristics of the circumbinary ring and that the binary is much closer to the north side of the circumbinary ring.

In addition, some have suggested that the gap in the northwestern edge of the circumbinary ring, which
can also be seen in Figure 1(a) and (b) as darker areas, is a shadow cast by some circumstellar materials \citep{Krist2005,Itoh2014}.
However, none of the inner disk structures discovered in this study appear to be responsible for it, 
and thus the origin of this gap remains unknown.

Both stars in the $PI$ image are surrounded by bright nebula-like structures with radii of approximately 0.14$\arcsec$ (20.0 AU) and 0.10$\arcsec$ (13 AU) for GG Tau Aa and
GG Tau Ab, respectively. These nebula-like structures look
like circumstellar disks around each stars, 
but they could also be dominated by speckles. To help distinguish between speckles and real disk
structures in the inner region, 
a vector map of the $PI$ image was constructed. 
The fact that some vectors show centrosymmetric characteristics surrounding the central stars implies the presence of real disk structures. 
It is much more difficult to judge what the noncentrosymmetric vectors imply at this stage; thus, we will leave discussions on such vectors to the future study.

The vector maps centered on GG Tau Aa with sizes of $4\arcsec\times4\arcsec$ and $1\arcsec\times1\arcsec$ are shown in Figures 2(a) and (b), respectively, where the white bars show the polarization angles (PAs). We calculated PAs $\theta_{p}$ using the formula $\theta_{\rm p}=0.5{\rm tan^{-1}}(U/Q)$
with binned data of 7 and 3 pixels for Figures 2(a) and (b), respectively. 
The errors of the PAs were estimated from the noise of the Stokes $Q$ and $U$ images. 
The typical error of the PAs was approximately $5^\circ$ for the circumbinary ring, and that for the inner region of the disk was approximately $3^\circ$. 
The vectors in the bright nebula-like structures around both stars do not show centrosymmetric characteristics, which may indicate that these bright nebular appearances are dominated by speckles.

However, the vectors to the north of GG Tau Ab show a region with
obvious centrosymmetry, extending approximately 0.40$\arcsec$ (56
AU) to the north of GG Tau Ab with a signal-to-noise ratio (SNR) of larger than approximately $5\sigma$.
The brightest part in this region could have an SNR of approximately $11\sigma$. 
Here, the SNRs of the $PI$ image were calculated from the SNRs of the Stokes $Q$ and $U$ images. 
Therefore, this area shows a real structure that may correspond to the north arc reported by \citet{Krist2002}.
We refer it hereafter as "the north arc". 
Its inner side appears to be close to 0.10$\arcsec$ from GG Tau Ab and to connect the two "bridges" mentioned by \citet{Itoh2014}, 
which are barely noticeable in this image. 
The eastern bridge has an SNR of approximately $4\sigma$, indicating it may be real. 
On the other hand, the western bridge has an SNR of only approximately $3\sigma$; therefore, the detection of this bridge remains uncertain. 
In the southern part of the inner disk, such obvious disk structures were not observed. 
This overall feature suggests that the inner region may be asymmetric.

The vectors outside the bright structures in GG Tau Aa, especially
vectors to the northeast of the star, tend to be centrosymmetric, which
indicates that part of the disk structures around GG Tau Aa were captured in this image. 
The outermost boundary appears to extend to approximately 0.20$\arcsec$ (28 AU) in projection.
This is slightly larger than but still in fair agreement with the radius of approximately 20 AU previously
reported for the circumstellar disk of GG Tau Aa (e.g., \citealp{Dutrey2014}). 
No such circumstellar
structures are discernible in the present image of GG Tau Ab.
Based on the speckle radius of GG Tau Ab, the radius of the disk structure around the GG Tau Ab1/Ab2 binary is constrained to < 13 AU.

The bright structure between the two stars indicated in Figure 2 appears to be connected, but the vectors are not centrosymmetric. 
Considering the possible complexity of the polarization pattern between two stars, it is still unclear whether this structure is real or simply speckles. 
Observations in other bands could be helpful in improving our understanding of this potential structure.

\section{Discussion}

\subsection{Binary Orbit}

According to \citet{Beust2005}, the semi-major axis (SMA) of the GG Tau A binary could be either 32 or 62 AU. 
Strict fitting of the astrometric data yielded an SMA of 32 AU. 
However, a binary with such a small SMA could not open such a large gap in the circumbinary disk. 
Another attempt taking into larger error bars of the astrometric data gave SMA = 62 AU that could fit the size of the ring but had a significance of only $3\sigma$. 
They concluded that the disk and the binary were likely to be coplanar but the astrometric data errors were underestimated. 
\citet{kohler2011} noted that not only the underestimation of astrometric data errors but also the misalignment of the binary orbit plane and disk could be responsible for the discrepancy between the astrometric data and the ring size.

We compared present binary position with the observation performed by \citet{Beck2012} in late 2009, approximately five years prior to the observation considered in this study.
The PA was found to have changed by approximately $7^{\circ}$ over that time. 
In the study by \citet{Beust2005}, both the 32- and 62-AU models have a PA rate of approximately $1.4^\circ/yr$, which is consistent with the present results. 
However, in this case, the SMA of the binary could not be constrained using only astrometry.
Thus, the binary orbit was constrained using the disk structure model. 
\citet{Pelupessy2013} developed a formula relating the radius of the density peak in the circumbinary disk to the SMA and eccentricity of the binary orbit:
\begin{equation}
a_{peak}=(3.2+2.8e_{binary})a_{binary},
\end{equation}
where $e_{binary}$ and $a_{binary}$ are the eccentricity and SMA of the binary, respectively, and $a_{peak}$ is the radius of the density peak in the disk.

First, we attempted to constrain the SMA from the surface brightness peak and the density peak locations.
For both $a_{binary}=32$ AU and $a_{binary}=62$ AU (with an eccentricity of $e=0.35$) the locations of the surface density peak
are 130 and 260 AU, respectvitly, using the equation by \citet{Pelupessy2013}.
Both numbers did not coincide with that of the surface brightness
peak at 180 AU.  
The surface brightness and surface density are
likely to peak at different radii because the former is sensitive primarily 
to the disk shape and less sensitive to the density. 
Therefore we concluded that the surface density peak lies outside the peak of the surface
brightness since scattering is likely dominated by the material closest to
the inner cavity where the illumination comes directly from the stars.
This favors the $a=62$ AU solution for the binary SMA.

The binary orbit could also affect the circumstellar disk around both stars. In the simulation performed by \citet{Regaly2011}, they determined that for $a_{binary}=40$ AU and $e=0.3$, the circumstellar disks around the companion stars should be approximately 13 AU. It is expected that if $a_{binary}=62$ AU, the disk radius should be larger than 13 AU, whereas if $a_{binary}=32$ AU, the disk radius is much smaller than 13 AU. 
In the simulation performed by \citet{Nelson2016}, they demonstrated that a binary with $a_{binary}=62$ AU and $e=0.3$ should have a circumstellar disk radius equal to 10 AU, whereas this radius is only 4 AU for $a_{binary}=32$ AU. In the present observation, a possible disk structure was detected around GG Tau Aa extending to a projected distance of approximately 28 AU, and \citet{Dutrey2014} noted that the circumstellar disk should have a radius on the order of approximately 20 AU. Both of these results suggest that GG Tau Aa has a relatively large disk. 
Although it has not been definitively determeined whether the circumbinary ring and the binary are misaligned, the results of the present study indicate that $a_{binary}=62$ AU is more likely than $a_{binary}=32$ AU.

\subsection{Structure in the Inner Region}

The observations made in this study provided the first high-resolution image of the inner region
around GG Tau A. 
Generally, it appears to be asymmetric. 
An arc structure was detected north of GG Tau Ab; however,
no such large disk-related feture was detected in the southern part. 
One would expect a binary comprising two stars with almost
the same mass have a symmetric disk structures. 
To better understand the structure in the inner region, we compare the present observations
with theory and previous observations, especially regarding CO gas and dust continuum
emissions.

\citet{Farris2014} calculated the accretion of binary black holes in circular orbits. Because the GG Tau A binary has a mass ratio $q$ of approximately 0.88, its accretion would be similar to the $q=0.82$ case shown in Figure 3 of \citet{Farris2014}. This figure shows that the streamers are asymmetric even though the binary orbit was circular in their simulation. They concluded that the asymmetry may be caused by the eccentric shape of the inner edge of a circumbinary disk driven by the binary tides, which lead to the different distances from the circumbinary disk to the binary in different directions. Such asymmetric streamers have also been described by \citet{Nelson2016}, who made simluation of the GG Tau A binary in an orbit with an eccentricity of 0.3. Considering that the north arc appears to connect the northern side of the ring and GG Tau Ab, the north arc may be part of a large streamer extending from the circumbinary ring to the inner disk. As mentioned in Section 3, the circumbinary ring is asymmetric, and the binary is much closer to the north; thus, the streamer from the north is larger than that in the south. 

In previous CO J=6--5 observations by \citet{Dutrey2014} with ALMA ,
two asymmetric CO cores were detected in the inner disk; implying possible interfaces of the streamers from the outer ring to the circumstellar disks around both stars. For a detailed comparison,
the CO 6--5 image is superimposed on the present $PI$ image in Figure 3(a). The north arc coincides with the position of the northern CO core observed by \citet{Dutrey2014}, suggesting that the north arc observed in this study could be part of a streamer observed in the NIR band.

For an advanced investigation, we checked the CO 6--5 velocity map in detail. 
The analysis of the CO 6--5 velocity field near the northern part of the CO 6--5 core of the inner region performed by
\citet{Dutrey2014} revealed a large velocity dispersion of approximately 2--2.5
km/s, which is larger than the predicted Keplerian rotation velocity dispersion of 1.2 km/s. 
This could be further evidence of the existence of the streamer.

We also compared the present observations with the H$_2$ emission observations
by \citet{Beck2012}, which revealed the temperature distribution in the inner disk. 
Figure 3(b) shows a peak in the H$_2$ emission represented by the darkest region that partly coincides with the north arc on its southern boundary, 
indicating a high temperature (approximately 2000 K) in this location. 
Such high temperatures tend to be the result of shockwaves in the inner disk, which are likely caused by inflows as suggested by \citet{Beck2012},
whereas the north arc observed in this study is slightly north of the peak of the H$_2$ emission. 
The peak of the hydrogen emission is located between the north arc and GG Tau Ab. 
One simple explanation for this is that the north arc observed in scattered light does not have a very high temperature 
and the temperature increases only when material begins to drop rapidly to GG Tau Ab.

The CO 6--5 core near the circumstellar disk around GG Tau Aa coincides with the possible disk structure. 
In the CO J=3--2 map presented by \citet{Tang2016}, there is one structure extending from GG Tau Aa; 
thus, part of this possible disk structure could also be part of a streamer feeding the circumstellar disk around GG Tau Aa. 
This may explain why it is slightly larger than the previously reported disk size. However, the CO 3--2 velocity map does not show clear sign of the material falling into GG Tau Aa, 
and the velocity field near GG Tau Aa from the CO 6--5 velocity map is too complex to draw a conclusion. 
Because the resolutions of the CO maps obtained by \citet{Tang2016} and \citet{Dutrey2014} are relatively low (approximately $0.3\arcsec$ and $0.25\arcsec$ for the CO 3--2 and 6--5 maps, respectively), 
a higher-resolution observation may aid the further analysis of the velocity field in this disk.

\subsection{Planet Formation}

Based on the parameters given in Table 3 of \citet{Andrews2014} and assuming a gas-to-dust ratio of 100:1, the Toomre Q parameter at disk radius 235 AU is $\sim5$.  Thus the ring is gravitationally stable, and a planet cannot form here through gravitational instability.

Previous CO (J=6--5, 3--2, and 2--1) images show a hotspot on the southwestern edge of the ring at a radius of approximately 250--260 AU
that has a temperature of about 40 K, which is 20 K higher than those in other locations at the same distance from the GG Tau A binary \citep{Dutrey2014, Tang2016}.  
It has been suggested that this hotspot is a signature of a potential planet. 
However, we see no corresponding structure in the present $PI$ image. 
The lack of such a structure could be due to a low degree of polarization. 
The mass of this CO hotspot reported by \citet{Tang2016} was only approximately $2M_{\rm J}$, thus it could be too faint to detect even in a NIR intensity image.

For the circumstellar disks around the binaries, if the mass of the outer disk ($0.15M_{\odot}$; \citealp{Dutrey1994})
and the total accretion rate of both stars ($5.1\times10^{-8} \; M_{\odot}/$yr; \citealp{Beck2012}) are taken in account,
the circumbinary gas rservoir can sustain the inner disk for at least 3 Myr. 
Because some planets with similar ages have been discovered so far, such as LkCa 15 b with an age of 2 Myr \citep{Kraus2012a, Sallum2015}, 
a duration of 3 Myr could be sufficient for a planet to form.

Radio continuum observations (e.g., \citet{Dutrey2014} and \citet{Tang2016}) have revealed the presence of large dust structures in the circumstellar disk around GG Tau Aa. 
Minimum disk mass estimate of GG Tau Aa is approximately $1M_{\rm J}$\citep{Dutrey2014}, which may not be enough for the formation of a Jupiter-like planet 
but may be feasible to form a Neptune-like or terrestrial mass planet \citep{Rafikov2015}.

The direct detection of the circumstellar disk around GG Tau Ab has not been reported yet.
Some studies such as that by \citet{Skemer2011}, and the presence of streamer to GG Tau Ab, give indirect evidence of the presence of the circumstellar disk.
Moreover, because GG Tau Ab itself is a binary, \citet{DiFolco2014} has noted that a putative disk associated with Ab would have been tidally truncated. As a result, the disk size around either GG Tau Ab1 or Ab2 must be less than 1/3 of the binary separation of GG Tau Ab, which is approximately 4.2 AU, and its circumbinary disk radius can be no larger than 13 AU.  Therefore, the environment
around GG Tau Ab binary may be hostile to planet formation.

\section{Conclusion}

Using Subaru/HiCIAO with the AO188 system,
a high-spatial-resolution (0.07$\arcsec$) image of the circumbinary disk
around the GG Tau A binary was successfully obtained. In comparison with previous observations,
the present polarimetry observations provide a much more detailed view of the disk structure inside the circumbinary ring. The present results indicate that the circumbinary disk around the binary is asymmetric and the binary is much closer to the northern edge of the ring than to the southern edge. By analyzing the sizes of the ring's inner edge and the circumstellar disk around GG Tau Aa, it was determined that the large semi major axis solution of 62 AU for the binary orbit is more likely than the small semi major axis solution of 32 AU.
An arc structure north of GG Tau Ab called the north arc in this paper
and a possible circumstellar disk structure around GG Tau Aa were observed inside the circumbinary ring. 
A comparison of the present observation results with previous observations and theoretical calculations suggests that
the north arc may be part of a large streamer extending from the circumbinary ring to GG Tau Ab. 
Based on previous estimates of the accretion rate and the outer disk mass, the streamer to the circumstellar
disk around each star may provide enough material for sub-Jovian planets to form in the disk around GG Tau Aa. 
Considering the circumstellar disk around each star, it seems that GG Tau Aa has a better chance to form a Neptune-like or terrestrial planet than GG Tau Ab.
This discovery may help reveal one aspect of the formation process of planets located in close binaries such as $\gamma$ Cep Ab, and it
may be helpful in improving our understanding of the planet formation process in binary star systems.

\acknowledgements{}

We would like to thank an anonymous reviewer, whose comments greatly helped us improve this paper. 
This study was based on data collected by Subaru Telescope, which is
operated by the National Astronomical Observatory of Japan (NAOJ), National Institutes of Natural Sciences (NINS). 
We thank the Subaru Telescope staff for their support during the observations. 
We would like to acknowledge the use of the SIMBAD database operated by the Strasbourg Astronomical Data Center (CDS), Strasbourg, France. This paper makes use of the following ALMA data: ADS/JAO.ALMA\#2011.0.00059.S. ALMA is a partnership of ESO (representing its member states), NSF (USA) and NINS (Japan), together with NRC (Canada), NSC and ASIAA (Taiwan), and KASI (Republic of Korea), in cooperation with the Republic of Chile. The Joint ALMA Observatory is operated by ESO, AUI/NRAO and NAOJ. 
IRAF is distributed by the National Optical Astronomy Observatory, which is operated by the Association of Universities for Research in Astronomy (AURA) under a cooperative agreement with the National Science Foundation. 
We also thank Dr. Ruobing Dong for useful discussions with him. MT is supported by a Grant-in-Aid for Scientific Research (No.15H02063). JC is supported by the U.S. National Science Foundation under Award No. 1009203.
The authors wish to recognize and acknowledge the very significant cultural role and reverence that the summit of Maunakea has always had within the indigenous Hawaiian community. We are most fortunate to have the opportunity to conduct observations from this mountain. 

\appendix
\section{APPENDIX}

Even with the AO system, residual seeing error still remains after correction.
This type of residual called a PSF halo, distorts the final
image and its polarization directions, as shown in Figure 6(a).
Therefore removal of PSF halo is necessary. 
The general process of removing the polarized halo which was not
fully corrected by the AO system involves rebuilding the
polarized halos for Stokes $Q$ and $U$ images then subtracting them
from the original images. After that, the corrected image
can be used to produce a halo-corrected polarized intensity image.

To create an artificial PSF halo around the binary,
the radial brightness profile of the two stars must first be calibrated in the Stokes $I$ image.
The radial profiles of the two stars can be derived using IRAF and Python scripts,
and the luminosity $L$ and radius $r$ are then fit by the function
\begin{equation}
L=A\,{\rm exp}(-r^{B}/C^{2}),
\end{equation}
where $A$, $B$, and $C$ are fitting parameters. The radial profile and
fitting results are shown in Figure 4(b) and (c). Considering that only the
polarized halos require fitting, the profiles are fit from 15 to
140 pixels to exclude the disturbances from the central Airy disk PSF.

After $A$, $B$, and $C$ have been determined for each halo, the observed
brightness ratio of the halo in the Stokes $Q$ and $U$ images must be compared
to that of the Stokes $I$ image to help regenerate the polarized halos in the Stokes $Q$
and $U$ images. For every Stokes image, photometry is conducted
by first calibrating the flux with apertures of 15 and 140 pixels
for GG Tau Aa and Ab, respectively.
Then subtracting the flux within 15 pixels from the flux calibrated within 140 pixels for each star. 
Thus, the flux of the PSF halo can be obtained for each star in all Stokes images. 
The ratio of the halo brightness in the Stokes $Q$ and $U$ images to that of the Stokes $I$
image is then obtained.
Based on the fitting and photometry results, the polarized halo for each star
can be regenerated in the Stokes $Q$ and $U$ images in the end.

For the case of a single star, the next step is to subtract the generated
PSF halo from the original Stokes $Q$ and $U$ images. 
However, in the case of a binary,
an extra step is required to combine the halos of the two companion
stars in the Stokes $Q$ and $U$ images. 
Considering that the halos are generated from
the observation results, the PSF halo from the star itself and the
effect of the other star are both included in the fitted halos.
That is why the two halos cannot be combined by simply adding them together.
Here, the maximum values of the two halos were mixed, i.e.,
for the halo $H_{a}$ around GG Tau Aa and that $H_{b}$ around GG Tau Ab, 
the final combined halo would be $H_{c}={\rm max}(H_{a},H_{b})$.
The combined halos for the Stokes $Q$ and $U$ images are shown in Figure 5. 
After these combined halos have been obtained, they are subtracted from each of the corresponding original Stokes images to obtain the halo-corrected $PI$
images (Figure 6(b)). In Figure 6(b), it can be seen that the vectors are generally centrosymmetric rather than aligned in one direction, which demonstrates that
the vectors were corrected successfully using this method.

\clearpage{}

\begin{figure*}[t]
\begin{centering}
\includegraphics[scale=0.2]{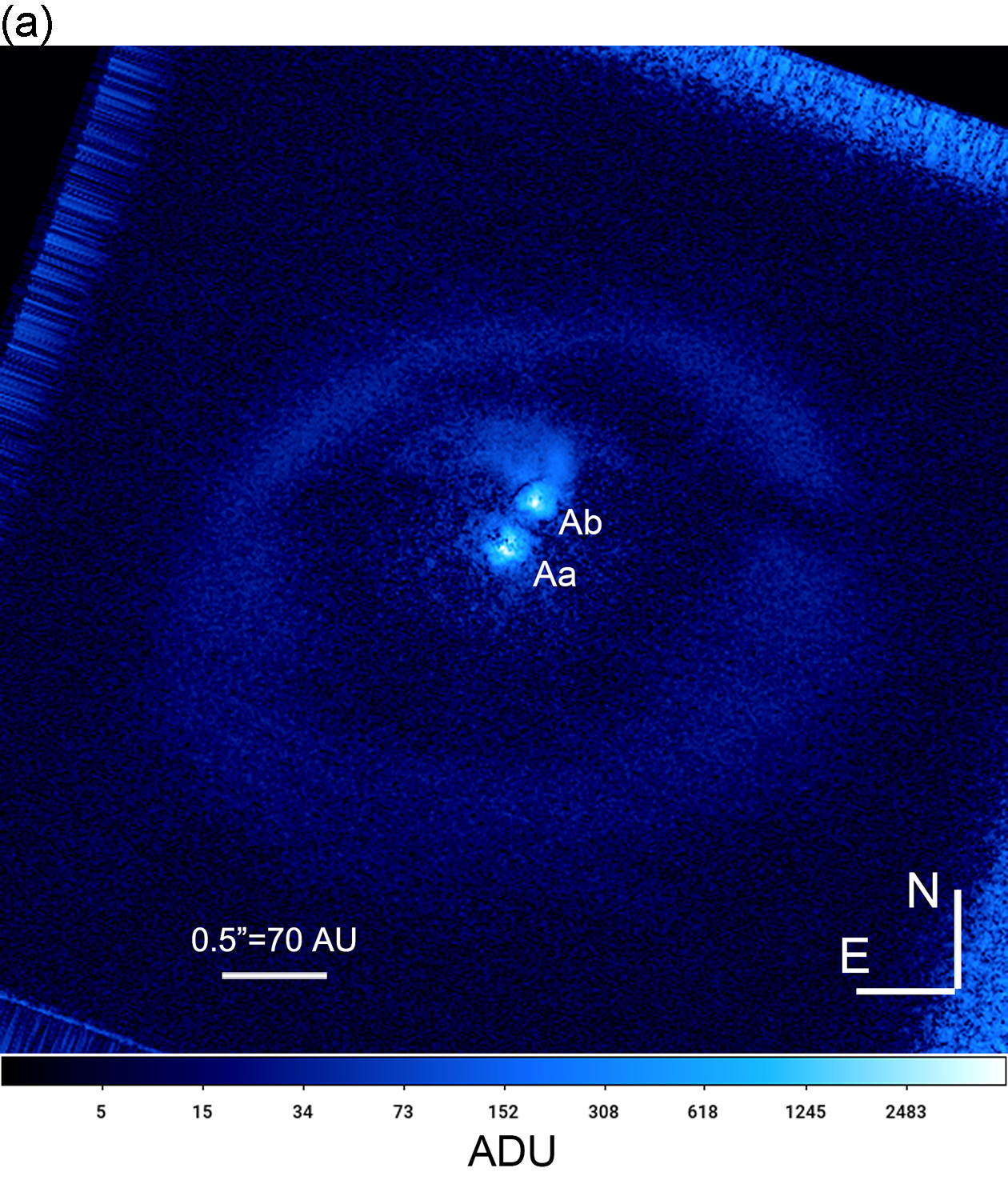}
\label{fig:a}
\includegraphics[scale=0.2]{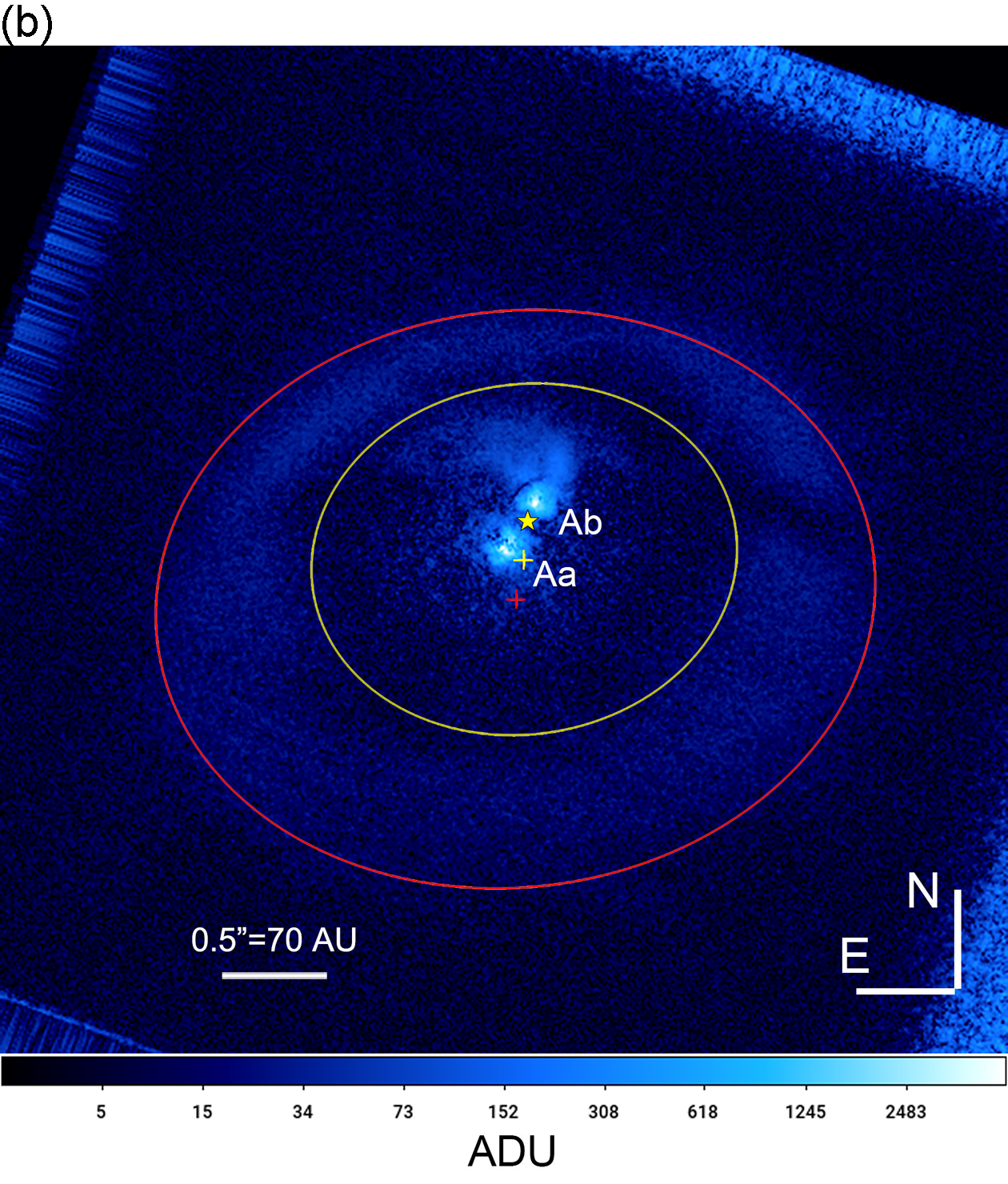}
\label{fig:b}
\par\end{centering}

Figure 1: (a): Polarized intensity image ($PI$ image) of GG Tau A ($512\times512$
pixels, corresponding to $4.9\arcsec\times4.9\arcsec$). The rectangle-like
structure around the image is an artifact caused by the data reduction process. (b): Comparison between the modeled circumbinary ring and the present observation. The inner and outer edges are shown in yellow and red, respectively. The yellow and red crosses represent the centers of the inner and outer edges, respectively. The yellow star represents one of the foci of the inner edge. Based on the theory, the barycenter of the binary should be near this focus. \end{figure*}

\begin{figure*}[t]
\begin{centering}
\includegraphics[scale=0.4]{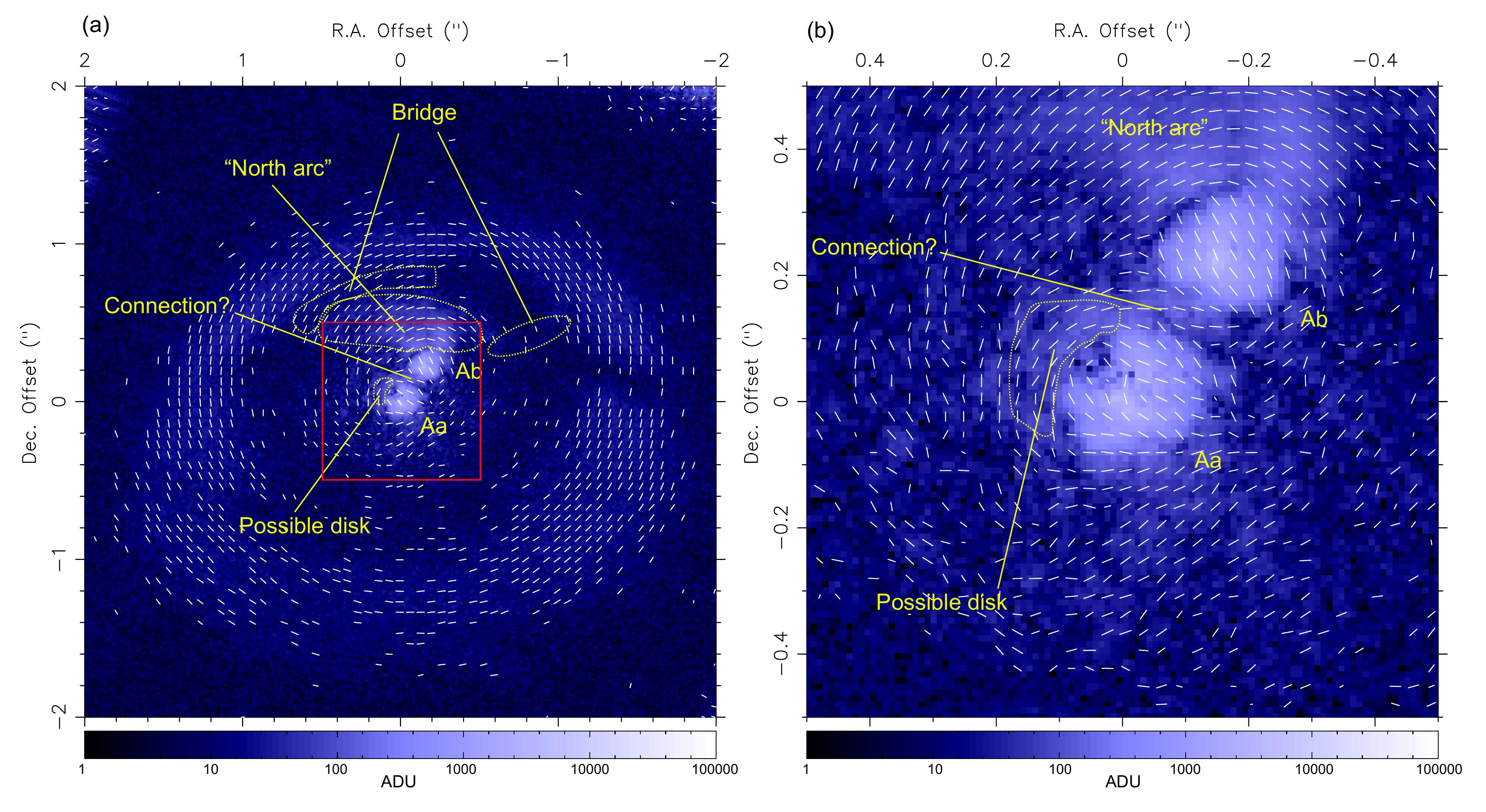}
\par\end{centering}

Figure 2: Polarization vector maps of GG Tau A near the central region
for the $PI$ image with fields of view of (a) $4\arcsec\times4\arcsec$ and (b) $1\arcsec\times1\arcsec$ centered on GG Tau A. (b) is an enlarged view of the area outlined by a red square in (a).
The polarization angles $\theta_{\rm p}$ were calculated using the formula
$\theta_{\rm p}=0.5{\rm tan}^{-1}(U/Q)$ in bins of (a) 7 and (b) 3 pixels,
and only areas brighter than 15 analog-to-digital unit (ADU) counts are drawn. The two bridges described by \citet{Itoh2014} are barely observable in this image. In the bright nebular-like structure
around GG Tau Aa and Ab, the vectors are clearly not centrosymmetric,
implying the presence of speckles. The north arc structure
to the north of GG Tau Ab and the structures to the northeast of GG
Tau Aa show centrosymmetric characteristics, representing possible real structures.
\end{figure*}

\begin{figure*}[t]
\begin{centering}
\includegraphics[scale=0.2]{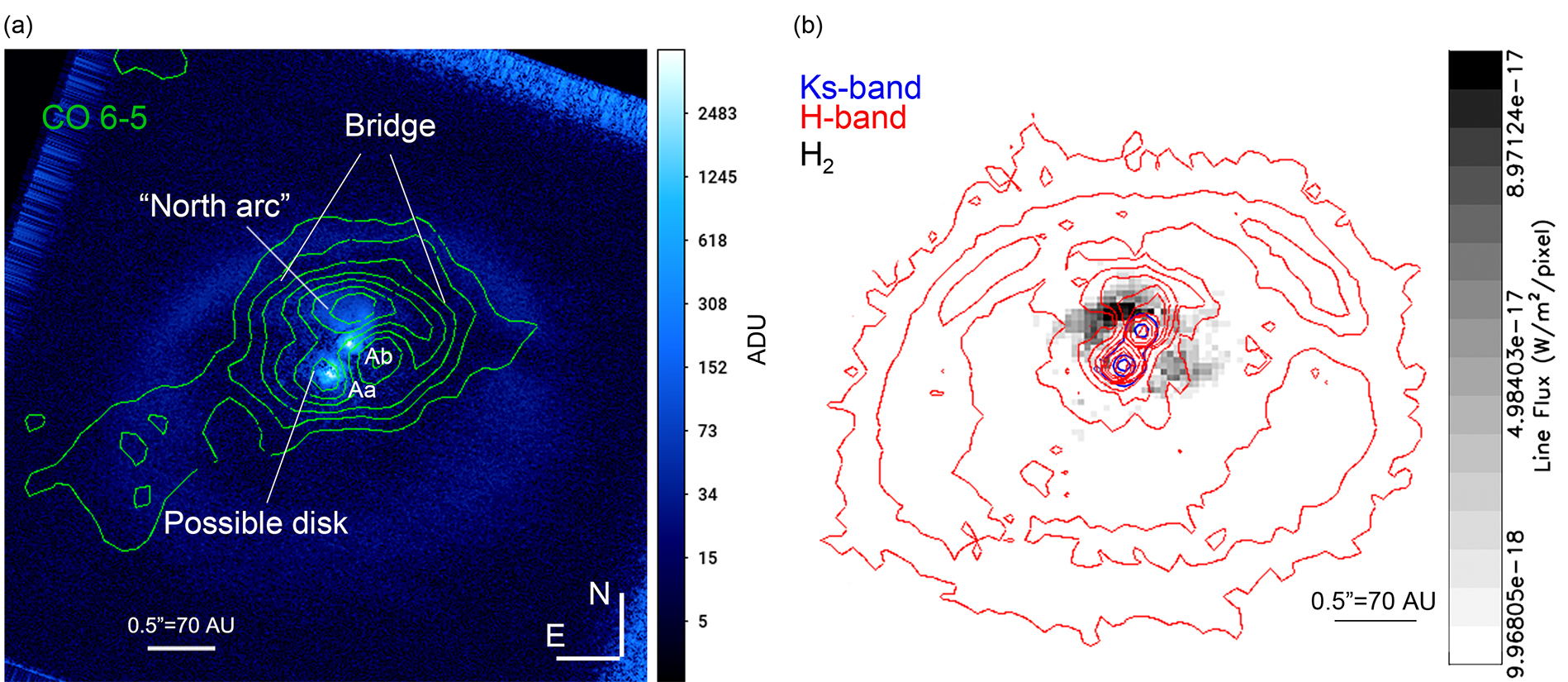}
\par\end{centering}

Figure 3: Comparison of present observations of the GG Tau disk with (a) previous ALMA CO 6--5
observations by \citet{Dutrey2014} and (b) H$_2$ emissions obtained by \citet{Beck2012}. The blue counters were obtained by \citet{Beck2012} and show the positions of GG Tau Aa and Ab.
\end{figure*}

\begin{figure*}[t]
\begin{centering}
\includegraphics[scale=0.7]{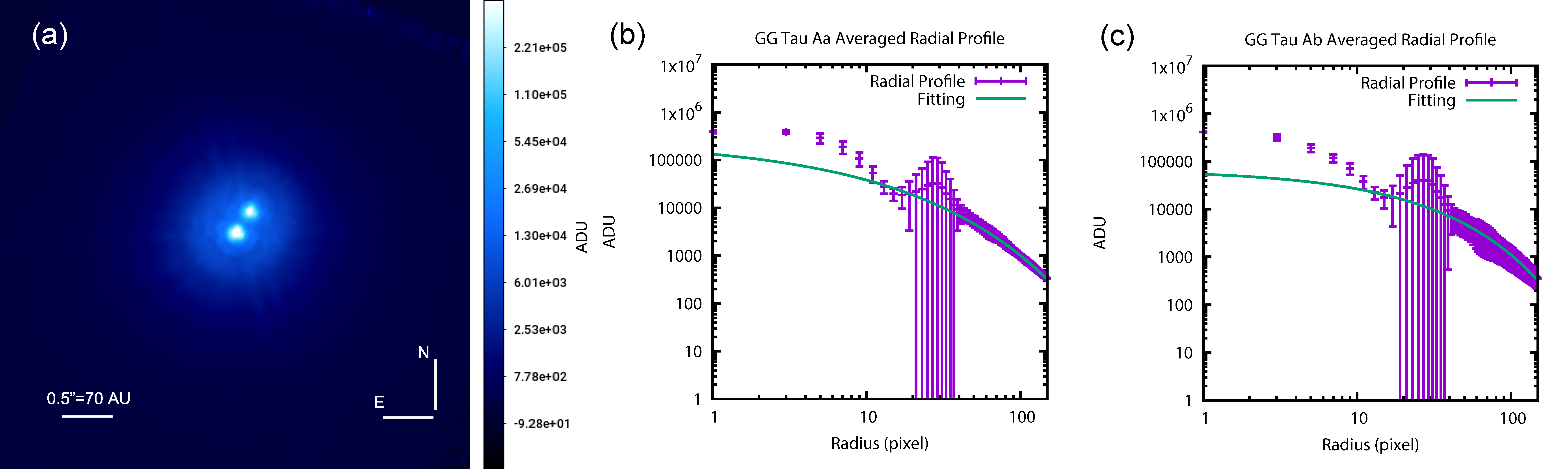}
\par\end{centering}

Figure 4: (a) Stokes $I$ image of GG Tau A. Radial profile (purple) and fitting results (green) for (b) GG Tau Aa
and (c) GG Tau Ab. The profiles were fit from 15 to 140 pixels so that only the PSF halo part was fitted. The large uncertainties at approximately 30 pixels were caused by the
other star and did not affect the fitting results.
\end{figure*}

\begin{figure*}[t]
\begin{centering}
\includegraphics[scale=0.3]{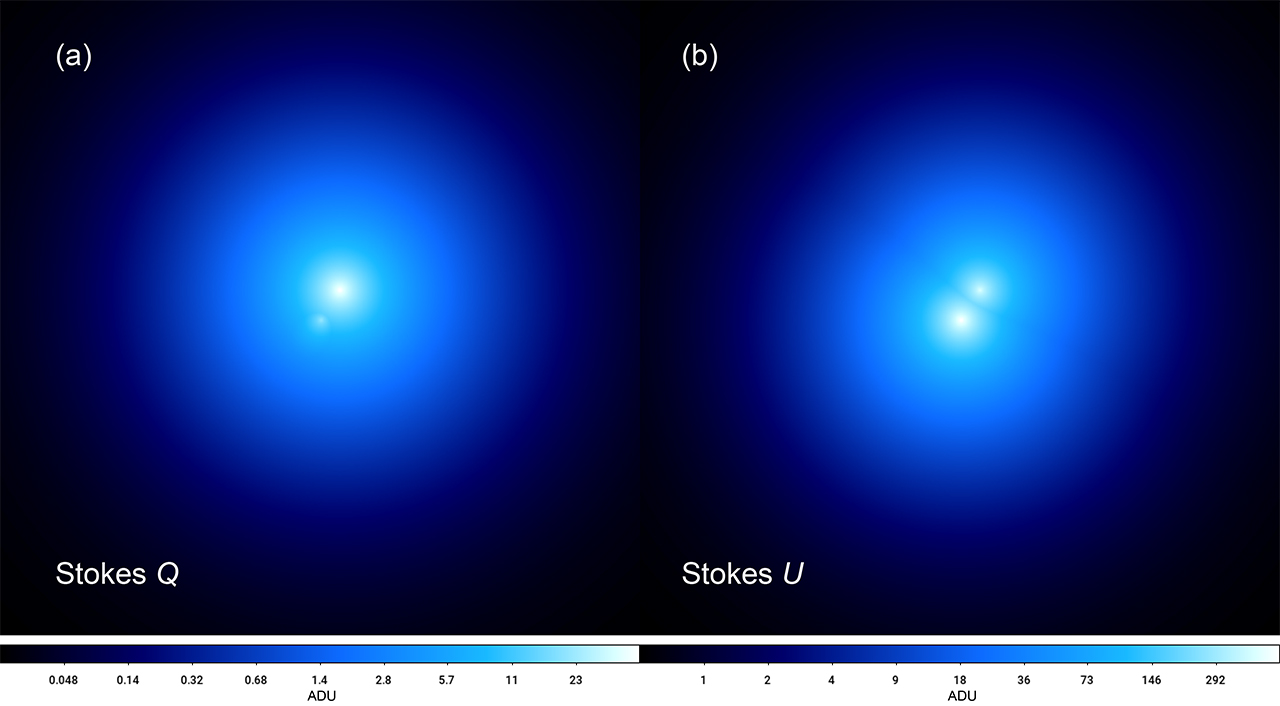}
\par\end{centering}

Figure 5: Generated PSF halos for (a) Stokes $Q$ and (b) $U$
images.
\end{figure*}

\begin{figure}
\begin{centering}
\includegraphics[scale=0.4]{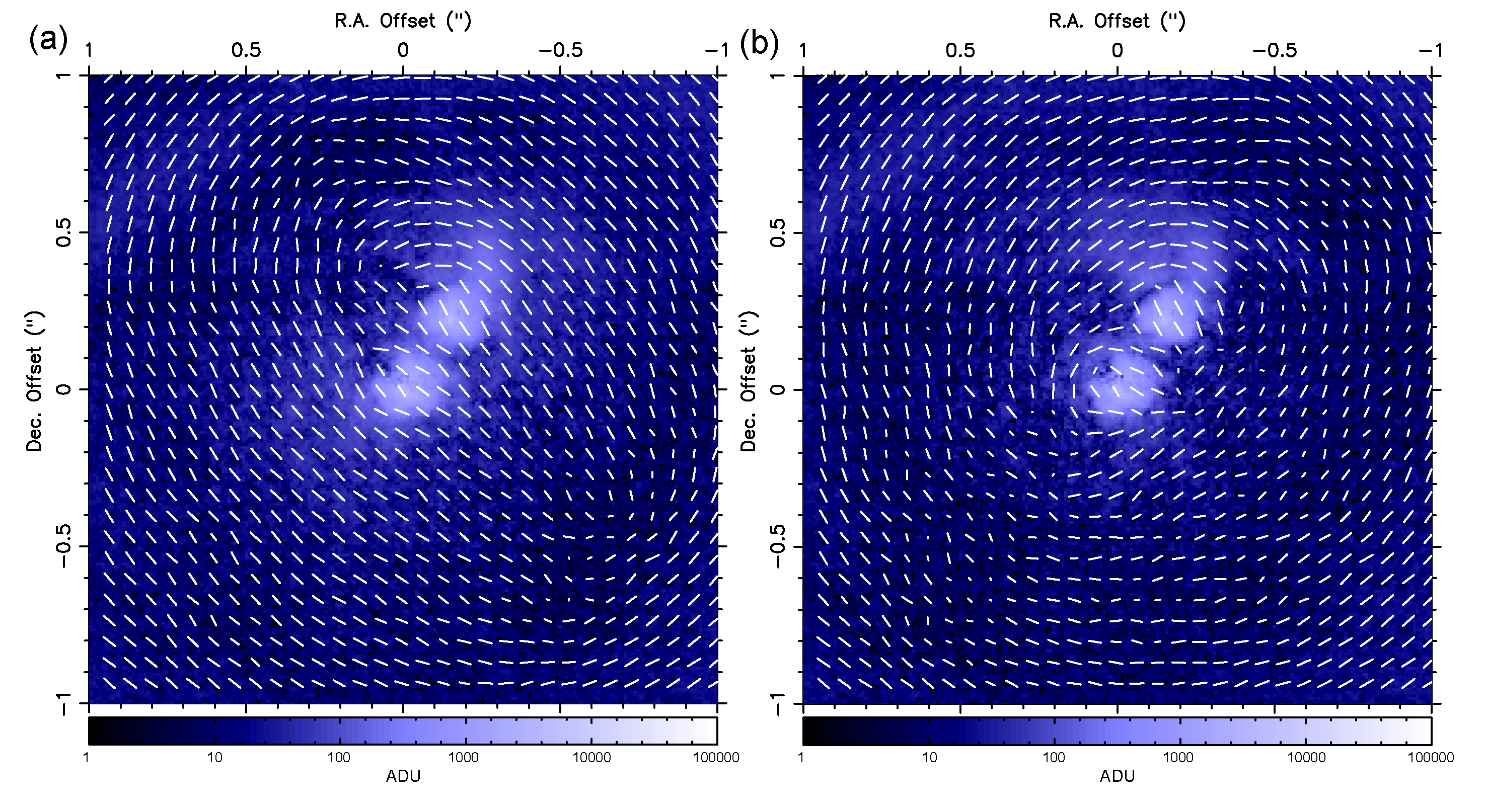}
\par\end{centering}
Figure 6: Vector map (a) before and (b) after PSF halo subtraction, showing the $2\arcsec\times2\arcsec$ area of GG Tau A and its disk.
\end{figure}

\listofchanges


\begin{thebibliography}{}
\bibitem[Andrews et al.(2014)]{Andrews2014}Andrews, S. M., Chandler, C. J., Isella, A., et al. 2014, \apj, 787, 148

\bibitem[Artymowicz \& Lubow(1994)]{Artymowicz1994}Artymowicz, P., \&
Lubow, S., 1994, \apj, 421, 651\textendash 667

\bibitem[Beck et al.(2012)]{Beck2012}Beck, T. L., Bary, J. S., Dutrey, A., et al. 2012,
\apj, 754, 72

\bibitem[Beust \&{} Dutrey(2005)]{Beust2005}Beust, H., \& Dutrey, A., 2005, \aap, 439, 585\textendash 594

\bibitem[Butler et al.(1997)]{Butler1997}Butler, R. P., Marcy, G. W., Williams, E.,
Hauser, H., \& Shirts, P., 1997, \apj, 474, L115\textendash L118

\bibitem[Currie et al.(2014)]{Currie2014}Currie, T., Daemgen, S., Debes, J.,
et al. 2014, \apj, 780, L30

\bibitem[Di Folco et al.(2014)]{DiFolco2014}Di Folco, E., Dutrey, A., Le
Bouquin, J., et al. 2014, \aap, 565, L2

\bibitem[Doyle et al.(2011)]{Doyle2011}Doyle, L., Carter, J., Fabrycky, D., et
al. 2011, Science, 333, 1602\textendash 1606

\bibitem[Duch{\^e}ne(2010)]{Duchene2010}Duch{\^e}ne G., 2010, \apj, 709, L114\textendash L118

\bibitem[Duch{\^e}ne \&{} Kraus(2013)]{Duchene2013}Duch{\^e}ne, G., \& Kraus, A.,
2013, \araa, 51, 269\textendash 310

\bibitem[Dutrey et al.(1994)]{Dutrey1994}Dutrey, A., Guilloteau, S., \& Simon, M.,
1994, \aap, 286, 149\textendash 159

\bibitem[Dutrey et al.(2014)]{Dutrey2014}Dutrey, A., Di Folco, E., Guilloteau, S.,
et al. 2014, Nature, 514, 600\textendash 602

\bibitem[Farris et al.(2014)]{Farris2014} Farris, B.~D., Duffell, P., MacFadyen, A.~I., \& Haiman, Z.\ 2014, \apj, 783, 134

\bibitem[Guilloteau et al.(1999)]{Guilloteau1999}Guilloteau, S., Dutrey, A.,
\& Simon, M., 1999, \aap, 348, 570\textendash 578

\bibitem[Hatzes et al.(2003)]{Hatzes2003}Hatzes, A. P., Cochran, W. D., Endl, M., et
al. 2003, ApJ, 599, 1383\textendash 1394

\bibitem[Hayano et al.(2010)]{Hayano2010}Hayano, Y., Takami, H., Oya, S., et al.\ 2010, \procspie, 7736, 77360N

\bibitem[Horch et al.(2014)]{Horch2014}Horch, E. P., Howell, S. B., Everett, M. E., \&
Ciardi, D. R., 2014, \apj, 795, 60

\bibitem[Itoh et al.(2014)]{Itoh2014}Itoh, Y., Oasa, Y., Kudo, T., et al. 2014,
RAA, 14, 1438\textendash 1446

\bibitem[Kenyon et al.(1994)]{Kenyon1994}Kenyon, S.J., Dobrzycka, D., \& Hartmann, L., 1994, \aj, 108, 1872

\bibitem[K{\"o}hler(2011)]{kohler2011}K{\"o}hler, R., 2011, \aap, 530, A126

\bibitem[Kozai(1962)]{Kozai1962}Kozai, Y., 1962, \apj, 67, 591

\bibitem[Kraus \&{} Hillenbrand(2009)]{Kraus2009}Kraus, A. L., \& Hillenbrand, L. A.,
2009, \apj, 704, 531\textendash 547

\bibitem[Kraus et al.(2012a)]{Kraus2012a}Kraus, A. L., \& Ireland, M. J., 2012,
\apj, 745, 5

\bibitem[Kraus et al.(2012b)]{Kraus2012b}Kraus, A. L., Ireland, M. J., Hillenbrand, L. A.
\& Martinache, F., 2012, \apj, 745, 19

\bibitem[Krist et al.(2005)]{Krist2005}Krist, J. E., Stapelfeldt, K. R., Golimowski, D. A.,
et al. 2005, \aj, 130, 2778\textendash 2787

\bibitem[Krist et al.(2002)]{Krist2002}Krist, J. E., Stapelfeldt, K. R., \& Watson, A. M.,
2002, \apj, 570, 785\textendash 792

\bibitem[Lidov(1962)]{Lidov1962} Lidov, M.~L.\ 1962, \planss, 9, 719

\bibitem[Mayama et al.(2010)]{Mayama2010}Mayama, S., Tamura, M., Hanawa, T.,
et al. 2010, Science, 327, 306\textendash 8

\bibitem[Nelson \& Marzari(2016)]{Nelson2016} Nelson, A.~F., \& Marzari, F.\ 2016, \apj, 827, 93

\bibitem[Palla \&{} Stahler(2002)]{Palla2002}Palla, F., \& Stahler, S. W., 2002,
\apj, 581, 1194\textendash 1203

\bibitem[Pelupessy \& Portegies Zwart(2013)]{Pelupessy2013} Pelupessy, F.~I., \& Portegies Zwart, S.\ 2013, \mnras, 429, 895
\bibitem[Perrin et al.(2009)]{Perrin2009} Perrin, M.~D., Schneider, G., Duchene, G., et al.\ 2009, \apjl, 707, L132

\bibitem[Pi{\'e}tu et al.(2011)]{Pietu2011}Pi{\'e}tu, V., Gueth, F., Hily-Blant, P.,
Schuster, K. \textendash F. \& Pety, J., 2011, \aap, 528, A81

\bibitem[Rafikov \& Silsbee(2015)]{Rafikov2015} Rafikov, R.~R., \& Silsbee, K.\ 2015, \apj, 798, 70

\bibitem[Raghavan et al.(2006)]{Raghavan2006}Raghavan, D., Henry, T. J., Mason, B. D.,
et al. 2006, \apj, 646, 523\textendash 542

\bibitem[Reg{\'a}ly et al.(2011)]{Regaly2011} Reg{\'a}ly, Z., S{\'a}ndor, Z., Dullemond, C.~P., \& Kiss, L.~L.\ 2011, \aap, 528, A93

\bibitem[Roddier et al.(1996)]{Roddier1996}Roddier, C., Roddier, F., Northcott, M. J.,
Graves, J. E., \& Jim K., 1996, \apj, 463, 326

\bibitem[Roell et al.(2012)]{Roell2012}Roell, T., Neuh{\"a}user, R., Seifahrt, A.,
\& Mugrauer, M., 2012, \aap, 542, A92

\bibitem[Sallum et al.(2015)]{Sallum2015}Sallum, S., Follette, K. B., Eisner, J. A., et al. 2015, \nat, 527, 342

\bibitem[Silber et al.(2000)]{Silber2000}Silber, J., Gledhill, T., Duchene, G.,
\& Menard, F., 2000, \apj, 536, L89\textendash L92

\bibitem[Skemer et al.(2011)]{Skemer2011}Skemer, A. J., Close, L. M., Greene, T. P., et
al. 2011, \apj, 740, 43

\bibitem[Tamura et al.(2006)]{Tamura2006}Tamura, M., Hodapp, K., Takami, H., et al.\ 2006, \procspie, 6269, 62690V

\bibitem[Tang et al.(2016)]{Tang2016}Tang, Y., Dutrey, A., Guilloteau. S., et
al. 2016, \apj, 820, 19

\bibitem[Wang et al.(2015)]{Wang2015} Wang, J., Fischer, D.~A., Xie, J.-W., \& Ciardi, D.~R.\ 2015, \apj, 813, 130

\end{thebibliography}
\end{document}